\documentclass[prl,reprint, twocolumn,showpacs,preprintnumbers,amsmath,amssymb,nofootinbib,floatfix]{revtex4-1} 
\usepackage{graphicx}

\usepackage{mathrsfs}
\usepackage{hyperref}

\usepackage{slashed}

\usepackage{color}

\usepackage{gensymb}

\usepackage{amsmath}
\allowdisplaybreaks[4]

\usepackage{bm}

\def \matrix #1 {\left(\begin{array}{cc} #1 \end{array}\right)}

\def\II{\hbox{{1}\kern-.25em\hbox{l}}}

\begin{document}

\title{Next-to-Next-to-Leading-Order QCD Prediction for the Pion  Form Factor }

\author{Yao Ji$^{a,b}$}
\email{corresponding author: yao.ji@tum.de}

\author{Bo-Xuan Shi$^{c}$}
\email{corresponding author: shibx@mail.nankai.edu.cn}

\author{Jian  Wang$^{d,e}$}
\email{corresponding author: j.wang@sdu.edu.cn}

\author{Ye-Fan  Wang$^{f}$}
\email{corresponding author: wangyefan@nnu.edu.cn}

\author{Yu-Ming Wang$^{c}$}
\email{corresponding author: wangyuming@nankai.edu.cn}

\author{Hui-Xin Yu$^{c}$}
\email{corresponding author: yuhuixin@mail.nankai.edu.cn}

\affiliation{\vspace{0.2 cm}
${}^a$ School of Science and Engineering, The Chinese University of Hong Kong,
Shenzhen, Guangdong, 518172,  China \\
${}^b$ Physik Department T31, James-Franck-Stra{\ss}e1, Technische Universit\"{a}t  M\"{u}nchen,
D–85748  Garching,  Germany \\
${}^c$ School of Physics, Nankai University, \\
Weijin Road 94, Tianjin 300071, P.R. China \\
${}^d$ School of Physics, Shandong University, Jinan, Shandong 250100, China \\
${}^e$ Center for High Energy Physics, Peking University, Beijing 100871, China \\
${}^f$  Department of Physics and Institute of Theoretical Physics,
Nanjing Normal University,  Nanjing, Jiangsu 210023, China
\vspace{0.2 cm}}

\date{\today}

\begin{abstract}
\noindent
We accomplish for the first time the two-loop computation of the leading-twist contribution
to the pion electromagnetic form factor by employing the effective field theory formalism rigorously.
The next-to-next-to-leading-order  short-distance matching coefficient is determined  by
evaluating the appropriate $5$-point QCD amplitude with the modern multi-loop technique
and subsequently  by  implementing  the ultraviolet renormalization and  infrared subtractions
with the inclusion of  evanescent operators.
The renormalization/factorization scale independence of the obtained  form factor
is then validated explicitly at ${\cal O}(\alpha_s^3)$.
The yielding two-loop QCD correction to this fundamental quantity turns out to be numerically significant
at experimentally accessible momentum transfers.
We further demonstrate that the newly computed two-loop radiative correction is highly  beneficial for
an improved determination of the leading-twist pion distribution amplitude.
\\[0.4em]

\end{abstract}

\preprint{TUM-HEP-1533/24}

\maketitle

%
\section{Introduction}
%

It is generally accepted that the pion electromagnetic form factor (EMFF)
is  of paramount importance for probing the infrared structure of QCD scattering amplitudes
of  hard exclusive reactions at leading power and beyond
and for unraveling the precise mechanisms that  dictate the intricate nature of composite hadron systems.
Advancing our understanding towards this gold-plated form factor is especially suitable for
delivering  deep and far-reaching insights into the delicate interplay
between the emergent and Higgs  mass generation mechanisms for such lightest pseudoscalar mesons
as the Nambu-Goldstone  modes of QCD.
Historically, the systematic analysis of the pion  form factor  played an indispensable role
in establishing the perturbative factorization formalism for the entire domain of exclusive hadronic processes
with large momentum transfers \cite{Lepage:1979zb,Lepage:1980fj,Efremov:1978rn,Efremov:1979qk,Duncan:1979hi,Duncan:1979ny,Brodsky:1989pv}.
As a consequence,  the dimensional-counting rules  for  the power-law behaviour of
numerous  hard  exclusive reactions on the basis of the parton model \cite{Brodsky:1973kr,Matveev:1973ra}
can be  placed on a solid footing within this field-theoretical framework  (see \cite{Gross:2022hyw} for an overview).
Additionally, the developed  perturbative factorization technique has been  extended to
the first-principles calculations of the deeply virtual Compton scattering \cite{Ji:1996nm,Radyushkin:1997ki,Diehl:1999tr,Diehl:1998kh},
the hard diffraction of light mesons at high energies \cite{Collins:1996fb,Brodsky:1994kf,Pire:2021hbl},
and the exclusive heavy-to-light $B$-meson decays at large hadronic recoil \cite{Beneke:2000wa,Beneke:2003pa,Bauer:2000yr,Bauer:2002aj}.
Moreover, the time-like scalar-current pion form factors are intimately connected with
the chirally-enhanced weak annihilation contributions to the flagship charmless two-body hadronic $B \to \pi \pi$ decays   \cite{Beneke:2001ev,Duraisamy:2010vhf,Lu:2022kos},
which provide an essential testing ground for CP violation in the Standard Model (SM)
with controllable theoretical uncertainties.

Experimentally,  the pion EMFF at low momentum transfers
($Q^2 \in [0.015, \, 0.253] \, {\rm GeV^2}$)
has been measured from the elastic scattering of high-energy pions off
the atomic electrons in a liquid-hydrogen target at Fermilab \cite{Dally:1981ur,Dally:1982zk}
and CERN \cite{Amendolia:1984nz,NA7:1986vav}.
The challenging measurement of the pion form factor at intermediate momentum transfers,
up to $Q^2 = 10 \, {\rm GeV^2}$, has been further carried out at Cornell \cite{Bebek:1974ww,Bebek:1976qm,Bebek:1977pe},
DESY \cite{Brauel:1979zk,Ackermann:1977rp} and JLab \cite{JeffersonLabFpi:2000nlc,JeffersonLabFpi:2007vir,JeffersonLabFpi-2:2006ysh,JeffersonLab:2008gyl,JeffersonLab:2008jve},
by exploiting the so-called Sullivan mechanism
and by extrapolating the determined longitudinal cross section at negative  values of
the Mandelstam variable $t$  to the pion pole at $t=m_{\pi}^2$ \cite{Sullivan:1971kd}.
Extracting the charged  pion form factor over a wide range of the moderate momentum transfers
($Q^2 \in [2.0, \, 6.0] \, {\rm GeV^2}$) can be also anticipated from the E12-06-101 experiment
with the upgraded JLab accelerator \cite{Arrington:2021alx}.
Accessing the space-like pion form factor  at yet higher momentum transfers
($10 \,  {\rm GeV^2} \leq Q^2 \leq 40 \,   {\rm GeV^2}$) with high precision
will be one of the  key  targets for the forthcoming EIC experiment at BNL \cite{AbdulKhalek:2021gbh}.

According to the  hard-collinear factorization theorem \cite{Lepage:1979zb,Lepage:1980fj,Efremov:1978rn,Efremov:1979qk,Duncan:1979hi,Duncan:1979ny},
the leading-power contribution to the pion EMFF in the large-momentum expansion
can be expressed in terms of the short-distance matching coefficient
and the twist-two light-cone distribution amplitude (LCDA) for the (anti)-collinear  pion state.
The  next-to-leading order (NLO) QCD correction to  the perturbatively calculable hard  function
had been  explicitly determined with  the diagrammatic factorization approach more than forty years ago \cite{Field:1981wx,Dittes:1981aw,Sarmadi:1982ig,Khalmuradov:1984ij,Braaten:1987yy}
(see \cite{Melic:1998qr} for further verification).
Including the complete NLO radiative correction in the factorization analysis of the pion EMFF
turns out to be extraordinarily advantageous to pin down the intrinsic theory uncertainty
from varying the renormalization and factorization scales
in comparison with the counterpart tree-level prediction \cite{Melic:1998qr}.
Accomplishing the next-to-next-to-leading-order (NNLO) computation of the short-distance coefficient function
rigorously is therefore in high demand, on the one hand,
for enriching and developing the perturbative factorization formalism at an unprecedented level,
and on the other hand, for improving further the resulting theory prediction of the charged pion form factor
in order to match the ever-growing precision of the dedicated  experimental measurements at JLab and EIC.
It is our primary objective to fill such an important and longstanding gap in this Letter,
by evaluating an appropriate $5$-point partonic matrix element at ${\cal O}(\alpha_s^3)$
with the contemporary advanced  multi-loop computational strategies
and by carrying out the ultraviolet (UV) renormalization and infrared (IR) subtractions
in  a meticulous  and factorization-compatible manner \cite{CCFJ:2024}.
Phenomenological implications of the thus determined two-loop QCD correction to the space-like pion form factor
at large momentum transfers will be then explored comprehensively,
by adopting four sample models for the leading-twist pion distribution amplitude at a low reference scale.

%
\section{General analysis}
%

We first lay out the theoretical framework for constructing the hard-collinear factorization formula
of the pion EMFF at leading power in an expansion in powers of $\Lambda_{\rm QCD}^2/Q^2$,
where $\Lambda_{\rm QCD}$ stands for the strong interaction scale.
Applying the general decomposition for the hadronic matrix element of the quark electromagnetic current
enables us to write down
\begin{eqnarray}
&& \langle \pi^{+}(p^{\prime}) | j_{\mu}^{\rm em}(0) |  \pi^{+}(p) \rangle
\nonumber \\
&& = F_{\pi}(Q^2) \, (p + p^{\prime})_{\mu} + \tilde{F}_{\pi}(Q^2) \, (p - p^{\prime})_{\mu}  \,,
\label{general decomposition of the hadronic matrix element}
\end{eqnarray}
where $p$ and $p^{\prime}$ correspond to the four-momenta carried by the initial and final pion states, respectively.
Employing the vector-current conservation condition immediately leads to $\tilde{F}_{\pi}(Q^2)=0$,
thus leaving us with a single form factor $F_{\pi}(Q^2)$.
Implementing further the charge conjugation transformation for the matrix element
on the left-hand side of (\ref{general decomposition of the hadronic matrix element})
indicates  an isospin  relation between  the two charged pion form factors
$F_{\pi^{-}}(Q^2) = - F_{\pi}(Q^2)$ for an arbitrary value of $Q^2$.
In  addition, the electric charge conservation determines the normalization condition
for the pion form factor in the forward-scattering limit $F_{\pi}(0)=1$
(see, for instance,  \cite{Khodjamirian:2020btr} for an elementary discussion).
Here the customary notation $Q^2= - (p-p^{\prime})^2$ has been employed
and the  quark electromagnetic current is given by
\begin{eqnarray}
j_{\mu}^{\rm em}(x)  =  \sum_{q} e_q \, \bar q(x) \gamma_{\mu} q(x) \,,
\label{definition of the EM current}
\end{eqnarray}
with $e_u=2/3$ and $e_d=-1/3$ for the up and down quarks.
Introducing two light-like reference vectors $n_{\mu}$ and $\bar n_{\mu}$
with the constraints $n^2=\bar n^2 =0$ and $n \cdot \bar n = 2$
then allows for  the decomposition $p_{\mu} =  ({n \cdot p} /2) \, \bar n_{\mu}$
and $p^{\prime}_{\mu} = ({\bar n \cdot p^{\prime}} /2) \, n_{\mu}$
at the leading-power accuracy.

Taking advantage of the modern effective theory field technique \cite{Bauer:2002nz,Rothstein:2003wh},
the hard-collinear factorization formula for the pion form factor at large momentum transfers
can be cast in the desired form (see \cite{Li:1992nu,Li:2010nn,Li:2012md,Li:2013xna,Li:2014xda} for an alternative formalism)
\begin{eqnarray}
F_{\pi}(Q^2) &=& (e_u - e_d) {4 \pi \alpha_s(\nu) \over Q^2}  f_{\pi}^2
\int d x \int d y \,  T_1(x, y, Q^2, \nu, \mu)
\nonumber \\
&& \times \, \phi_{\pi}(x, \mu)  \,\,  \phi_{\pi}(y, \mu) \,,
\label{factorization formula}
\end{eqnarray}
which is valid to all orders in perturbation theory and at leading power in the $\Lambda_{\rm QCD}^2/Q^2$ expansion.
We take the charged pion decay constant from the three-flavour  FLAG  average
$f_{\pi} = (130.2 \pm 1.2) \, {\rm MeV}$ \cite{FlavourLatticeAveragingGroupFLAG:2021npn}
with an increased uncertainty from adding  an approximate  $0.7 \%$  charm sea-quark contribution.
Apparently, the hard-scattering kernel $T_1$ depends on both the renormalization scale $\nu$
and the factorization scale $\mu$,  the latter of which corresponds to the resolution
with which the microscopic structure of the $\pi$-meson is being probed.
This short-distance coefficient function can be expanded perturbatively
in terms of the strong coupling constant (similarly for any other QCD quantity)
\begin{eqnarray}
T_1 = \sum_{\ell=0}^{\infty} \, \left ( {\alpha_s \over 4 \pi} \right )^{\ell} \, T_1^{(\ell)} \,.
\label{expansion of hard kernel}
\end{eqnarray}
The leading-twist pion distribution amplitude $\phi_{\pi}$ in the factorized expression (\ref{factorization formula})
can be defined by the renormalized QCD matrix element on the light-cone \cite{Braun:1988qv,Braun:1989iv}
\begin{eqnarray}
&& \langle  \pi^{+}(p^{\prime}) | \bar u (\tau \, \bar n) \,\, [\tau \, \bar n, 0] \,\, \gamma_\mu \,  \gamma_5 \, d(0) | 0 \rangle
\nonumber \\
&& = -i \, f_\pi \, p^{\prime}_\mu \, \int^1_0 dx \, e^{i x \, \tau \,  \bar n \cdot p^{\prime}}  \,
\phi_\pi(x, \mu) \,,
\label{definition: twist-2 pion DA}
\end{eqnarray}
where $[\tau \, \bar n, 0]$ is the  finite-length collinear Wilson line ensuring  gauge invariance.
The one-loop renormalization-group (RG) equation for the light-ray operator
on the left-hand side of (\ref{definition: twist-2 pion DA}) implies
the conformal partial wave expansion of $\phi_\pi(x, \mu)$ in terms of the Gegenbauer polynomials
with multiplicatively renormalizable coefficients (see \cite{Braun:2003rp} for an excellent overview)
\begin{eqnarray}
\phi_\pi(x, \mu)  = 6 \, x \, (1-x) \, \sum_{m=0}^{\infty} \, a_m(\mu) \,  C_{m}^{3/2}(2 x -1) \,.
\label{Gegenbauer expansion of pion LCDA}
\end{eqnarray}
The normalization condition $\int_0^1 d x \,\phi_\pi(x, \mu) = 1$ has been adopted throughout this work
and the odd moments $a_{1,3,...}(\mu)$ vanish due to the ${\rm G}$-parity symmetry.

The short-distance matching coefficient $T_1$ can be routinely determined
by  investigating the $5$-point QCD matrix element below
\begin{eqnarray}
\Pi_{\mu} = \langle u(p_1^{\prime}) \,  \bar d (p_2^{\prime}) |j_{\mu}^{\rm em}(0)
|  u(p_1) \,  \bar d (p_2)  \rangle   \,,
\label{definition: QCD amplitude}
\end{eqnarray}
where the external momenta can be restricted to their leading components
$p_1=x\, p$, $p_2=  \bar x \, p$, $p_1^{\prime}=y\, p^{\prime}$ and
$p_2^{\prime}=\bar y\, p^{\prime}$
with the  ``bar notation" $\bar x \equiv 1-x$,  $\bar y \equiv 1-y$.
We will perform the computation of the bare  amplitude
in dimensional regularization with $D=4 - 2 \, \epsilon$,
where UV and IR divergences manifest themselves  as poles up to the second order in $\epsilon$.
The former divergences are evidently cancelled by the UV renormalization for the strong coupling constant
$\alpha_s$ in the $\overline{\rm MS}$ scheme,
while the latter disappear after executing the  nowadays standard  IR subtraction procedure.

%
\section{Next-to-next-to-leading-order  QCD computation}
%

\begin{figure}[htp]
\includegraphics[width=0.45 \textwidth]{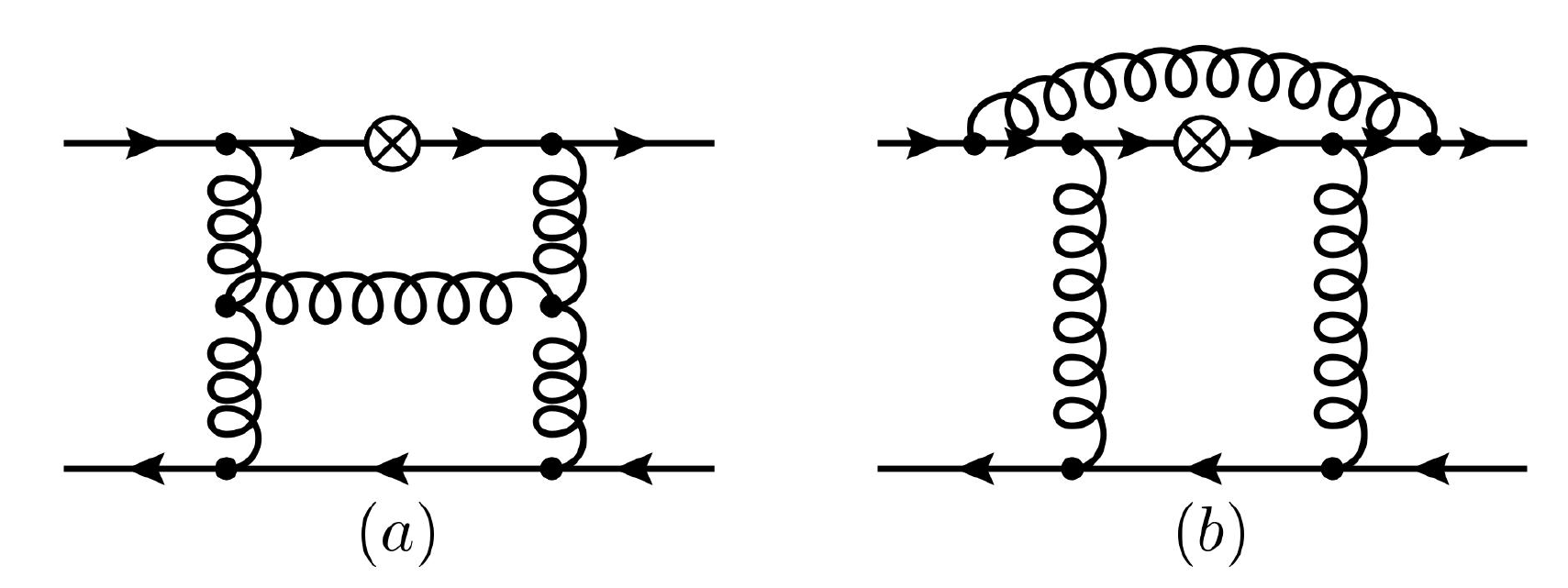}
\caption{Sample two-loop Feynman diagrams.
The circled cross $\otimes$ marks an electromagnetic current insertion. }
\label{fig: sample 2-loop diagrams}
\end{figure}

We will dedicate this section to a brief technical  description of the  two-loop QCD calculation
for the considered partonic quantity $\Pi_{\mu}$.
We start with generating   the NNLO Feynman diagrams  with  {\tt FeynArts} \cite{Hahn:2000kx}
and,  independently,   by means of an in-house routine.
Taking into account the observation that a subset of the two-loop diagrams
yield the vanishing contribution, due to the Furry theorem and/or  the zero colour (electric) charge factors,
we eventually encounter $1066$ non-vanishing diagrams of our interest.
The two sample Feynman diagrams are depicted in Figure \ref{fig: sample 2-loop diagrams} explicitly.

All tensor integrals can be expressed in terms of scalar integrals
and  algebraic tensorial structures using the Passarino-Veltman decomposition \cite{Passarino:1978jh}.
We then perform the Dirac and tensor reductions with in-house {\tt Mathematica} routines,
by applying  further the QCD equations of motion and on-shell conditions.
The large amount of two-loop scalar integrals are further reduced to
an irreducible set of master integrals with the  packages
{\tt Apart} \cite{Feng:2012iq} and {\tt FIRE} \cite{Smirnov:2008iw}.
In total, we obtain  $57$  two-loop  master integrals that appear in our computation,
$11/16/30$ of which appear to  depend on $1/2/3$  physical scale(s).
Unsurprisingly, we  encounter the entire list of the $12$ one- and two-scale  master integrals
entering  the NNLO computation of the photon-pion form factor \cite{Gao:2021iqq,Gao:2021beo}.
The evaluation of the remaining master integrals can be accomplished in an analytic fashion
by employing the method of differential equations \cite{Kotikov:1990kg,Gehrmann:1999as}.
In order to identify a suitable linear transformation such that the master integrals in the new basis
satisfy a set of differential equations in an $\epsilon$-factorized form (i.e., canonical form) \cite{Henn:2013pwa},
we take advantage of  Lee's algorithm \cite{Lee:2014ioa} as implemented in the program {\tt Libra} \cite{Lee:2020zfb}.
The boundary conditions to the   differential equations are determined
by  capitalizing on  the package  {\tt AMFlow} \cite{Liu:2022chg} based upon \cite{Liu:2017jxz,Liu:2021wks}.
The yielding numerical results of  the master integrals at three  distinct kinematic points
$( x, y ) \in \{ (1/2, 1/5), \, (1/3, 1/6), (1/4, 1/8)  \} $, with approximately $100$ digits,
allow us to construct the  desired boundary constants of the differential equations
in terms of transcendental numbers using the {\tt PSLQ} algorithm \cite{ferguson1999analysis,ferguson1998polynomial}.
It is thus straightforward to express the master integrals  through Goncharov polylogarithms (GPLs) \cite{Goncharov:1998kja,Goncharov:2001iea} up to the required order in $\epsilon$.
We will present the analytic expressions for all master integrals in the forthcoming write-up.

%
\section{Ultraviolet renormalization and infrared subtractions}
%

We proceed to deduce  the master formula for the hard-scattering kernel $T_1$
by performing the UV renormalization and IR subtractions in the presence of evanescent operators,
whose significance in the perturbative treatment of quantum field theory
has been explored at length ever since the pioneering era of dimensional regularization  \cite{Breitenlohner:1977hr,Bonneau:1980zp,Bonneau:1979jx,Collins:1984xc}.
To achieve this goal, we exploit the matching equation for the interested QCD amplitude $\Pi_{\mu}$
onto the effective  matrix elements
\begin{eqnarray}
\Pi_{\mu} &=& (p + p^{\prime})_{\mu} \, \left [  (e_u - e_d) \,  {4 \pi \alpha_s  \over Q^4}  \,
\sum_k \, T_k    \otimes \langle  {\cal O}_k  \rangle \right ],
\nonumber \\
{\cal O}_k & \in &  \left \{ {\cal O}_1, \,  {\cal O}_2, \, {\cal O}_3 \right \}  \,,
\label{matching equation}
\end{eqnarray}
with the following  choice of  the collinear operator basis
\begin{align}
{\cal O}_1 &= \left [ \bar \chi_{u} \slashed{\bar n}  \gamma_5   \chi_d  \right ]
\left [ \bar \xi_d   \slashed{n} \gamma_5   \xi_u \right ],
\nonumber \\
{\cal O}_2  &= \left [ \bar \chi_{u}  \gamma_{\perp \alpha}    \xi_u  \right ]
\left [ \bar \xi_d  \gamma_{\perp}^{\alpha}    \chi_d \right ] - {1 \over 4} \, {\cal O}_1,
\nonumber \\
{\cal O}_3  &= \left [ \bar \chi_{u}  \gamma_{\perp \alpha}   \gamma_{\perp \mu_1}  \gamma_{\perp \mu_2}  \xi_u  \right ]
\left [ \bar \xi_d  \gamma_{\perp}^{\alpha} \gamma_{\perp}^{\mu_2}   \gamma_{\perp}^{\mu_1}    \chi_d \right ].
\label{collinear operator basis}
\end{align}
One can readily verify that  ${\cal O}_1$ is the only physical operator in our problem
and the other two operators  are evanescent, i.e., vanish algebraically in four dimensions.
To reduce our notation to the essentials, we strip off the collinear Wilson lines
and the position arguments of quark fields from ${\cal O}_k$,
and represent them  merely  by their flavour and Dirac structure.
Following the  conventions of \cite{Beneke:2005vv},
we label the collinear and anti-collinear effective fields moving into the directions of
$\bar n$ and $n$ by $\xi$ and  $\chi$, respectively,
which evidently satisfy $\slashed{\bar n} \, \xi = 0$ and $\bar \chi \, \slashed{n}=0$.

We are now in a position to organize the perturbative expansion of the renormalized QCD correlation function $\Pi_{\mu}$
in terms of the partonic tree-level matrix elements of the effective operators ${\cal O}_k$
\begin{eqnarray}
\Pi_{\mu}  &=& (p + p^{\prime})_{\mu} \,\, (e_u - e_d) \,  {(4 \pi)^2  \over Q^4}  \,
\sum_k \, \sum_{\ell=1, 2, 3} \,
\nonumber \\
&&   \left   [   \left ( {Z_{\alpha}  \alpha_s \over 4  \pi} \right )^{\ell + 1} \,
A_{k}^{(\ell)} \otimes \langle  {\cal O}_k  \rangle^{(0)} \right   ]  \,,
\label{general QCD amplitude}
\end{eqnarray}
where the renormalization constant of the strong coupling in the standard $\overline{\rm MS}$ scheme
takes the form of $Z_{\alpha} = 1 -   {\alpha_s \over 4 \pi} \, {1 \over \epsilon} \, \beta_0 \,
+ \left ({\alpha_s \over 4 \pi} \right )^2  \left ( {1 \over \epsilon^2} \, \beta_0^2 - {1 \over 2 \epsilon} \, \beta_1 \right )
+ {\cal O}(\alpha_s^3)$
(see \cite{Herzog:2017ohr,Luthe:2017ttg,Chetyrkin:2017bjc} for the five-loop expression).
The scalar quantities $A_{k}^{(\ell)}$ in (\ref{general QCD amplitude}) represent the bare $\ell$-loop on-shell  QCD amplitudes.
In the same vein, the UV-renormalized matrix elements of the collinear operators ${\cal O}_k$
can be  expanded  according to
\begin{eqnarray}
\langle  {\cal O}_k  \rangle = \sum_{i} \, \sum_{\ell=0}^{\infty} \,
\left ( {\alpha_s \over 4 \pi} \right )^{\ell} \,
Z_{k i}^{(\ell)} \otimes \langle  {\cal O}_i  \rangle^{(0)} \,,
\label{expansion of collinear operators}
\end{eqnarray}
further  taking into account that  scaleless integrals vanish in dimensional regularization.
Since the collinear fields for distinct directions already decouple at the hard scale,
the renormalization constant $Z_{11}^{(\ell)}$ can be therefore determined from
the  celebrated Efremov-Radyushkin-Brodske-Lepage (ERBL) kernel with the one- and two-loop results
obtained in \cite{Lepage:1980fj, Efremov:1979qk} and \cite{Sarmadi:1982yg,Dittes:1983dy,Katz:1984gf,Mikhailov:1984ii,Belitsky:1999gu}.
Substituting Eqs. (\ref{expansion of hard kernel}),  (\ref{general QCD amplitude}) and  (\ref{expansion of collinear operators})
into the matching equation  (\ref{matching equation}) leads us to
derive the master formulae for the short-distance coefficient function
\begin{eqnarray}
T_1^{(0)} &=& A_1^{(0)} \,,
\nonumber \\
T_1^{(1)} &=&  A_1^{(1)} + Z_{\alpha}^{(1)} \, A_1^{(0)} - \sum_{k=1, 2} \, Z_{k 1}^{(1)} \otimes T_k^{(0)} \,,
\nonumber  \\
T_1^{(2)} &=&   A_1^{(2)} + 2 \, Z_{\alpha}^{(1)} \, A_1^{(1)}  + Z_{\alpha}^{(2)} \, A_1^{(0)}
\nonumber \\
&& - \sum_{k=1, 2, 3} \, \sum_{\ell =1, 2} \, Z_{k 1}^{(\ell)} \otimes T_k^{(2-\ell)}  \,,
\end{eqnarray}
by comparing the coefficient of the tree-level matrix element of the physical operator  $\langle  {\cal O}_1  \rangle^{(0)}$.
Following the prescription proposed in \cite{Dugan:1990df,Herrlich:1994kh,Buras:1989xd,Jamin:1994sv,Buras:2011we},
the renormalization constants for the evanescent operators are adjusted to ensure that
the IR-finite matrix elements $\langle  {\cal O}_{k}  \rangle$ ($k=2, 3$)
vanish at an arbitrary scale.
Adopting the preferred  collinear operator basis (\ref{collinear operator basis})
yields the peculiar  evanescent-to-physical operator  mixing under the RG evolution,
which turns out to be adequately captured by the {\it finite} renormalization constant $Z_{21}^{(2)}$ at two-loop order.
This justifies  the essential role of introducing evanescent operators in constructing the QCD factorization formulae
with  dimensional regularization beyond the leading logarithmic approximation \cite{Beneke:2005vv,Beneke:2006mk,Becher:2004kk,Hill:2004if,Beneke:2005gs,
Buchalla:1995vs,Buras:2000if,Wang:2015vgv,Wang:2017ijn,Gao:2019lta,Gao:2021iqq,Huang:2024ugd,Li:2020rcg,Cui:2023jiw}.

Inserting the newly obtained two-loop expression  of  the hard-scattering kernel $T_1$
into the collinear factorization formula (\ref{factorization formula})
and then performing the two-fold integration  over $x \in (0, 1)$ and $y \in (0, 1)$  analytically
with  the {\tt Mathematica} package {\tt PolyLogTools} \cite{Duhr:2019tlz} in the  asymptotic approximation
(namely, $\phi_\pi^{\rm Asy}(x, \mu)=6 \, x \, (1-x)$) results in
\begin{widetext}
\begin{eqnarray}
F_{\pi}^{\rm Asy} (Q^2)  &=& (e_u - e_d) \, {4 \pi \alpha_s(\nu) \over Q^2}
\,  9 \,  f_{\pi}^2  \, \left ( {C_F \over 2  N_c} \right )
\, \bigg \{  1 + {\color{magenta} \left( {\alpha_s  \over 4 \pi} \right )} \,
{\color{blue}   \left [ \beta_0 \, \ln{\nu^2 \over Q^2}
+  {14 \over 3} \, \beta_0   - {71 \over 6} \, C_F  +  {1 \over 3 N_c} \right ]  }
\nonumber \\
&& +  \, {\color{magenta} \left ( {\alpha_s \over 4 \pi} \right )^2 } \,
\bigg [ \left ( \beta_1 \, \ln{\nu^2 \over Q^2}  - \beta_0^2 \,  \ln^2{\nu^2 \over Q^2}  \right )
+ 2 \, \beta_0 \, \ln{\nu^2 \over Q^2} \,   {\color{blue} \left ( \beta_0 \, \ln{\nu^2 \over Q^2}
+ {14 \over 3} \, \beta_0   - {71 \over 6} \, C_F  +  {1 \over 3 N_c}  \right )}
\nonumber \\
&& + \, 4  \left ( C_F \, \beta_0 \, \left ( {5 \over 2}  - \zeta_2 \right )
-  2 \, C_F^2  (\zeta_2 + \zeta_3 )\right )  \, \ln{\mu^2 \over Q^2}
+ C_A^2 \left ( {34873 \over 81}  + {88 \over 3} \, \zeta_2  + {152 \over 3} \, \zeta_3 - 160 \, \zeta_5  \right )
\nonumber  \\
&& - \, C_A  \, C_F  \left ( {8191 \over 18}  + {1163 \over 9} \, \zeta_2 + 418 \, \zeta_3
- 2 \, \zeta_4 - 760 \, \zeta_5  \right )
+  \, C_F^2  \left ( 194  + 61 \, \zeta_2 + 246 \, \zeta_3  - 18 \, \zeta_4 - 560 \, \zeta_5  \right )
\nonumber \\
&& -  \, C_A \, n_{\ell} \, T_F  \left (  {21742 \over 81} +  {32 \over 3} \, \zeta_2
-  48 \, \zeta_3  + {160 \over 3} \, \zeta_5  \right )
+   C_F \, n_{\ell} \, T_F \left (  {769 \over 9} + {316 \over 9} \, \zeta_2 - 8 \, \zeta_3   \right )
+  \left (n_{\ell} \, T_F \right )^2   {3496 \over 81}    \bigg  ] \bigg \},
\hspace{0.80 cm}
\end{eqnarray}
\end{widetext}
where $C_F=(N_c^2-1)/(2 N_c)$ and $C_A=N_c$ denote the Casimir operators
of the fundamental and adjoint representations of the ${\rm SU}(N_c)$ gauge group
with the standard normalization $T_F=1/2$.
In addition,  $n_{\ell}$ stands for the number of active quark flavours,
and $\zeta_n$ represents  the Riemann zeta function with $\zeta_2= \pi^2/6$,
$\zeta_3 \cong 1.202056903$, $\zeta_4= \pi^4/90$,
and $\zeta_5 \cong 1.036927755$ \cite{Vermaseren:1998uu,Smirnov:2012gma}.
Including the higher conformal spin contributions from  the twist-two pion LCDA  $\phi_\pi$
leads to the lengthy results for the non-asymptotic corrections to the charged pion form factor,
whose explicit expressions with the $m=12$ truncation of the Gegenbauer expansion
are collected in the Supplemental Material for completeness.
In particular, we have  verified that the thus achieved NNLO QCD computation of $F_{\pi}(Q^2)$
with the perturbative factorization formula (\ref{factorization formula})
is truly independent of the renormalization/factorization scale
at the ${\cal O}(\alpha_s^3)$ accuracy, by applying the two-loop  evolution equation of $\phi_\pi(x, \mu)$  \cite{Sarmadi:1982yg,Dittes:1983dy,Katz:1984gf,Mikhailov:1984ii,Belitsky:1999gu}.
Taking advantage of the momentum-space RG formalism  enables us to accomplish an all-order summation of
the enhanced logarithms of $Q^2 / \Lambda_{\rm QCD}^2$ entering the factorized expression of $F_{\pi}(Q^2)$
in the next-to-next-to-leading-logarithmic (NNLL) approximation,
which necessitates an implementation of the three-loop evolution of
the leading-twist pion distribution amplitude \cite{Braun:2017cih,Strohmaier:2018tjo}.

%
\section{Numerical analysis}
%

We are now prepared to explore the phenomenological implication of the newly determined
two-loop  QCD correction to the pion EMFF,
with an emphasis on the detailed comparison with both the available  experimental measurements
and the state-of-the-art lattice QCD results.
To achieve this goal, we  proceed by first discussing our choice for  the phenomenological  models
of the  twist-two pion LCDA appearing in the hard-collinear factorization formula (\ref{factorization formula}).
The first model $\phi_{\pi}^{\rm Model \,\, I}(x, \mu_0) = \left [ \Gamma(2 + 2   \alpha_{\pi}) / \Gamma^2 (1 + \alpha_{\pi}) \right ] \,
(x \bar x)^{\alpha_{\pi}}$ is motivated from the anti-de Sitter-QCD correspondence \cite{Brodsky:2007hb}
with the non-perturbative parameter $\alpha_{\pi}(\mu_0)=0.585^{+0.061}_{-0.055}$ \cite{Khodjamirian:2020mlb}
determined by matching to  the updated  lattice result of
the  second Gegenbauer moment  $a_2(\mu_0)=0.116^{+0.019}_{-0.020}$
at the reference  scale $\mu_0= 2.0  \, {\rm GeV}$ \cite{RQCD:2019osh}.
Our  second model $\{a_2, \, a_4, a_6, a_8  \} \, (\mu_0)
= \{0.181(32), \, 0.107(36),  \, 0.073(50), \, 0.022(55) \}$ \cite{Cheng:2020vwr}
is obtained from  the comparison of the light-cone sum rule for the pion EMFF \cite{Braun:1999uj}
with the experimental measurements (see \cite{Bijnens:2002mg,Khodjamirian:2011ub} for the earlier construction along this line).
By contrast,  the numerical intervals for the two lowest conformal coefficients  in model III
$\{a_2, \, a_4 \}\, (\mu_0) =  \{0.149^{+0.052}_{-0.043}, \, -0.096^{+0.063}_{-0.058}  \}$  \cite{Bakulev:2001pa,Mikhailov:2016klg,Stefanis:2020rnd}
are extracted from the method of QCD sum rules with non-local condensates \cite{Mikhailov:1991pt}.
Furthermore, we employ an alternative model (hereafter labelled as model IV) with
the first three non-vanishing Gegenbauer coefficients
$\{a_2, \, a_4, \, a_6 \}\, (\mu_0) =  \{0.196(32), \,  0.085(26), \, 0.056(15)  \}$ \cite{Cloet:2024vbv}
determined from the  lattice  prediction for the  dynamic shape  of the pion distribution amplitude
in the range of $x \in [0.25, 0.75]$ using large momentum effective theory \cite{Ji:2013dva,Ji:2014gla,Ji:2020ect}
and from  a simple power-law parametrization of  the corresponding  end-point behaviour.
Additionally, we will vary the renormalization scale  for the strong coupling $\alpha_s$
in the interval $\nu^2 \in [Q^2/2,  \, 2 \, Q^2]$  with the central value $Q^2$.
Following \cite{Agaev:2010aq}, the factorization scale $\mu$ characterizing the virtualities of quark and gluon propagators
in the hard-scattering partonic process will be taken as $\mu^2 = \langle x \rangle \, Q^2$
with $1/4 \leq \langle x \rangle \leq 3/4 $ (see \cite{Melic:1998qr,Melic:2001wb,Stefanis:1997zyh,Mojaza:2012mf,He:2006ud,Lu:2009cm,Wang:2015ndk,Wang:2017ijn,DiGiustino:2023jiq}
for  an elaborate discussion on the renormalization/factorization scale setting).

\begin{figure}[htp]
\includegraphics[width=0.45 \textwidth]{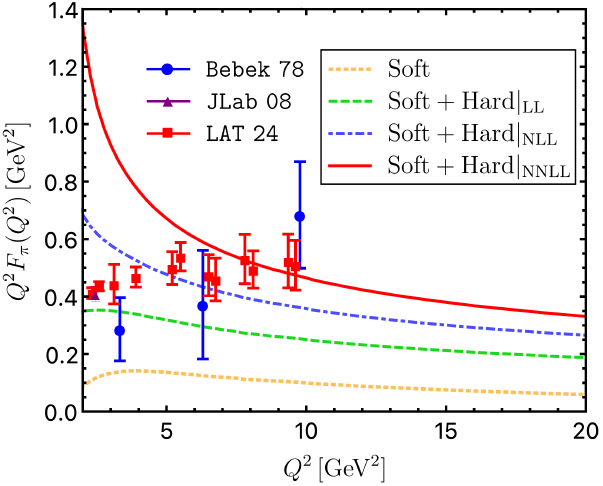}
\caption{Theory predictions for  the pion EMFF from
the soft ``end-point"  Feynman mechanism \cite{Braun:1999uj} (yellow curve)
and the (competing)  hard-scattering contribution
in the kinematic region $Q^2 \in [2.0, \, 20.0] \, {\rm GeV}^2$.
We further collect here  the experimental data points for the  space-like pion form factor
at intermediate momentum transfers
({\tt Bebek} 78 \cite{Bebek:1977pe} and {\tt JLab 08} \cite{JeffersonLab:2008jve})
and the most recent lattice QCD predictions with the physical pion  mass at
$Q^2=\{2.34, \, 2.58, \, 3.12, \, 3.90, \, 5.20, \,
5.50, \, 6.50, \, 6.75, \, 7.80, \, 8.10, \, 9.35,  \\
\,\, 9.61 \} \,\,  {\rm GeV^2}$({\tt LAT 24} \cite{Ding:2024lfj})
for an exploratory comparison.}
\label{fig: distinct contributions to the pion form factor}
\end{figure}

In order to facilitate  an in-depth exploration  of the dynamical pattern  dictating the pion form factor,
we  display  explicitly in Figure \ref{fig: distinct contributions to the pion form factor}
the  obtained theory predictions for the leading-power hard-gluon-exchange contributions
at leading-logarithmic (LL), NLL, and NNLL accuracy accomplished in this Letter
as well as the subleading power corrections arising from
I) an  atypical  ``end-point"  configuration of twist-two  in the one-loop approximation
and II) the higher-twist pion distribution amplitudes up to the twist-six accuracy with the dispersion technique \cite{Braun:1999uj},
by adopting model I of the pion distribution amplitude $\phi_{\pi}$ as our default choice.
Inspecting the distinctive feature for a variety of  higher-order corrections
in Figure \ref{fig: distinct contributions to the pion form factor} implies that
the newly determined two-loop QCD correction to the hard-scattering contribution
based upon the hard-collinear factorization theorem
can substantially enhance the corresponding  NLL prediction of the pion form factor
at intermediate and large momentum transfers: numerically at the level of $(30-50) \%$.
In analogy to the QCD anatomy of the radiative leptonic $\bar B \to \gamma \ell \bar \nu_{\ell}$ decay
with an energetic photon \cite{Braun:2012kp,Wang:2016qii,Wang:2018wfj,Beneke:2018wjp,Khodjamirian:2023wol},
the soft non-factorizable correction to the charged pion form factor
can constantly   shift the NNLL resummation improved leading-power contribution by
approximately an amount of ${\cal O} (25 \%)$ in the kinematic domain $5.0 \, {\rm GeV^2} \leq Q^2 \leq 20.0 \, {\rm GeV^2}$.
We are then led to conclude that an ironclad and {\it fully analytical} extraction of
the two-loop short-distance coefficient function $T_1^{(2)}$ in the factorized expression (\ref{factorization formula})
is evidently  vital for  obtaining  the robust and accurate theory prediction of the flagship hadron form factor $F_{\pi}$
and for advancing further the QCD factorization programme targeting at the high-precision computation of hard exclusive reactions.
It remains interesting to observe  that the very inclusion of the NNLO QCD correction to the pion EMFF
turns out to be especially beneficial for better accommodating the  benchmark lattice simulation results
at intermediate momentum transfers \cite{Ding:2024lfj}.

\begin{figure}[htp]
\includegraphics[width=0.45 \textwidth]{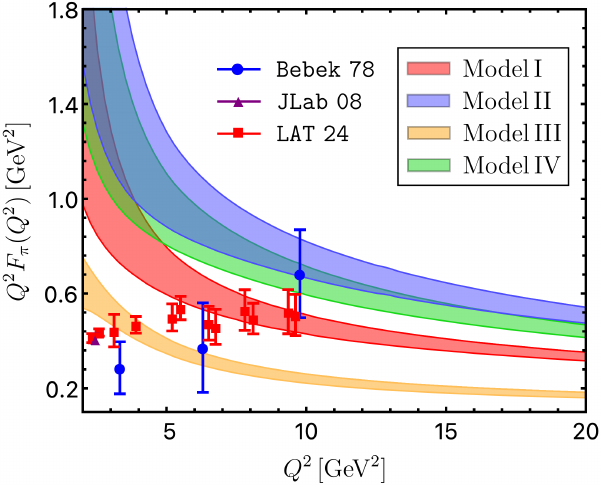}
\caption{Theory predictions for the pion EMFF  with  the four sample  models
of the leading-twist pion distribution amplitude $\phi_\pi(x, \mu)$
in the kinematic range $Q^2 \in [2.0, \, 20.0] \, {\rm GeV}^2$
obtained by adding together both the factorizable hard-gluon-exchange contribution at two loops
and the various power-suppressed contributions  discussed in the text.
We also display here the perturbative uncertainties from varying the renormalization and factorization scales
in the default intervals as indicated by the colour bands. }
\label{fig: the pion form factor with four models}
\end{figure}

We now turn to present in Figure \ref{fig: the pion form factor with four models} our final theory predictions
for the pion form factor $F_{\pi}$ with the four different phenomenological models
of the twist-two pion LCDA discussed above,
confronting with the two experimental measurements
{\tt Bebek} 78 \cite{Bebek:1977pe} and {\tt JLab 08} \cite{JeffersonLab:2008jve}
at four distinct kinematic points of $Q^2= \{2.45, \,  3.33, \,  6.30, \, 9.77 \}  \, {\rm GeV^2}$.
It can then be observed  that  the yielding numerical prediction with the anti-de Sitter-QCD inspired model
fulfilling an additional constraint from the lattice determination of $a_2(\mu_0)$ \cite{RQCD:2019osh}
(i.e., our model I) provides us with the most optimized description of  the available  experimental data points
and the model-independent lattice QCD results \cite{Ding:2024lfj} simultaneously.
The extraordinary snapshot of the well-separated uncertainty bands for the first three sample models
due to the variations of the renormalization/factorization scales in the preferred intervals
is  particularly encouraging to acquire new insights on the intricate behaviour of
the leading-twist pion distribution amplitude, in combination with the envisaged precision measurements at JLab and EIC.
In contrast with the NNLL QCD computation for the photon-pion transition form factor \cite{Gao:2021iqq,Braun:2021grd},
our two-loop prediction of the pion EMFF based upon the hard-collinear factorization prescription
approaches the desired scaling behaviour in the formal $Q^2 \rightarrow \infty$ limit rather slowly.
This intriguing pattern can be attributed to the fact that the determined hierarchy  between
the  subleading conformal spin  effect and the counterpart asymptotic contribution
grows steadily with the increasing loop order $\ell$ at realistic momentum transfers
accessible in the current and forthcoming experimental facilities,
thus postponing the onset of the asymptotic regime to an enormously higher value of $Q^2$
(see \cite{Braun:2001tj,Braun:2006hz,Anikin:2013aka,Huang:2024ugd}
for further discussions in the context of the nucleon electromagnetic form factors).

\begin{figure}[htp]
\includegraphics[width=0.45 \textwidth]{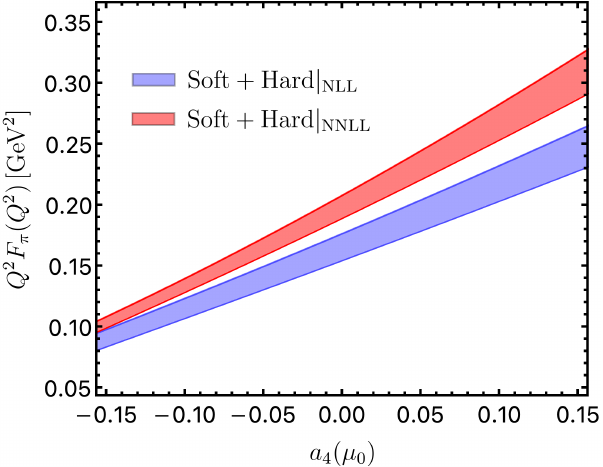}
\caption{Theory predictions for the dynamical dependence of the pion EMFF
on the fourth conformal coefficient $a_4(\mu_0)$ of the leading-twist $\pi$-meson distribution amplitude
at the sample kinematic point $Q^2=30.0 \,  {\rm GeV}^2$,
by taking advantage of the lattice determination of the second Gegenbauer moment
from the {\tt RQCD} Collaboration \cite{RQCD:2019osh}
and by further discarding the higher conformal moments
(namely, $a_{n \geq 6}(\mu_0)=0$) for illustration purposes.}
\label{fig: extraction of the fourth moment}
\end{figure}

We finally address the genuine benefit of  the  full two-loop QCD calculation
of the charged pion form factor on the model-independent extraction of the shape parameters
dictating the twist-two pion distribution amplitude on the light-cone.
To this end, we proceed to explore the intrinsic sensitivity of
the exclusive $\pi^{+} \, \gamma^{\ast}  \to \pi^{+}$ form factor
on the currently poorly constrained  conformal coefficient  $a_4(\mu_0)$
by combining the achieved NNLL  prediction of the leading-twist contribution
with the previously determined subleading power corrections in the $\Lambda_{\rm QCD}^2/Q^2$ expansion.
It is evident from Figure \ref{fig: extraction of the fourth moment} that
taking into account the newly obtained two-loop QCD correction to the pion EMFF form factor
will indeed be  advantageous to  enhance the sensitivity of extracting the key shape parameter $a_4(\mu_0)$
in anticipation of  the prospective precision EIC measurements \cite{AbdulKhalek:2021gbh},
complementary to an alternative strategy on the basis of
the factorization analysis of the photon-pion form factor \cite{Gao:2021iqq}.

%
\section{Conclusions}
%

In conclusion, we have endeavored to accomplish for the first time the rigorous two-loop QCD computation
of the pion form factor in an analytical fashion,
by applying the modern effective field theory formalism  that enables us to implement
the UV renormalization and IR subtractions of evanescent operators systematically.
Crucially, we have demonstrated that the thus determined NNLO QCD correction to the short-distance coefficient function
can bring about an enormous  impact on the pion form factor over a wide range of momentum transfers,
by employing the four phenomenologically acceptable models of the leading-twist pion distribution amplitude.
In particular, the very inclusion of the two-loop radiative correction allowed for an improved extraction
of the essential shape parameters dictating the intricate profile of the pion distribution amplitude,
when confronted with the encouraging measurements from the upcoming EIC experiment.
Extending and developing further our factorization prescription to the charged  kaon form factor
and to the more challenging  $B_c \to \eta_c \, \ell \, \bar \nu_{\ell}$ transition form factors
\cite{Boer:2018mgl,Bell:2005gw,Bell:2006tz,Boer:2023tcs}
will be highly beneficial for deepening our understanding towards the diverse facets of the strong interaction dynamics
encoded in the hard-scattering processes.

%
\begin{acknowledgments}
\section*{Acknowledgements}

It is our pleasure to thank  Yong-Kang Huang for illuminating discussions.
The research of Y.J. was supported in part by the Deutsche Forschungsgemeinschaft (DFG, German Research Foundation)
through the Sino-German Collaborative Research Center TRR110 ``Symmetries and the Emergence of Structure in QCD''
(DFG Project-ID 196253076, NSFC Grant No. 12070131001, - TRR 110).
J.W. is  supported   by the  National Natural Science Foundation of China
with Grant No. 12005117, No. 12321005, and No. 12375076,
and the Taishan Scholar Foundation of Shandong province with Grant No. TSQN201909011.
The work of Y.F.W. is supported in part by the  National Natural Science Foundation of China
with Grant No.  12405117.
Y.M.W. acknowledges support from the  National Natural Science Foundation of China
with Grant No.  12075125 and   No. 2475097.

\end{acknowledgments}

%
\appendix

\begin{widetext}

\section{SUPPLEMENTAL MATERIAL}

Here we collect  the general  expression for the derived two-loop  pion  electromagnetic form factor
with the inclusion of the non-asymptotic corrections
\begin{eqnarray}
F_{\pi}  (Q^2) &=& (e_u - e_d) \, {4 \pi \alpha_s(\nu) \over Q^2}
\,  9 \,  f_{\pi}^2  \, \left ( {C_F \over 2  N_c} \right ) \,
\sum_{m=0, 2, 4, ...}  \,\,  \sum_{n=0, 2, 4, ...}  \, a_{m}(\mu) \,\, a_{n}(\mu) \,
\nonumber \\
&& \times \, \bigg \{ 1  +  {\color{magenta}  \left ( {\alpha_s \over 4 \pi}  \right ) }\,
{\color{blue}   \left [ \beta_0 \, \ln{\nu^2 \over Q^2}
+ \left ( \gamma_{m, m}^{(0)} + \gamma_{n, n}^{(0)}   \right )  \, \ln{\mu^2 \over Q^2}
+{\cal R}_{m n}^{(1)} \right ] }
\nonumber \\
&& \hspace{0.5 cm} +  \,
{\color{magenta} \left ( {\alpha_s \over 4 \pi} \right )^2 } \,
\bigg  [ \left ( \beta_1 \, \ln{\nu^2 \over Q^2}  - \beta_0^2 \,  \ln^2{\nu^2 \over Q^2}  \right )
+ \, 2 \, \beta_0  \, \ln{\nu^2 \over Q^2} \, {\color{blue}   \left (   \beta_0 \, \ln{\nu^2 \over Q^2}
+ \left ( \gamma_{m, m}^{(0)} + \gamma_{n, n}^{(0)}   \right )  \, \ln{\mu^2 \over Q^2}
+ {\cal R}_{m,  n}^{(1)}   \right )  }
\nonumber \\
&& \hspace{0.5 cm} + \, \left ( \sum_{m^{\prime}=m}^{\infty}  \gamma_{m^{\prime}, m}^{(1)}
+ \sum_{n^{\prime}=n}^{\infty}  \gamma_{n^{\prime}, n}^{(1)} \right ) \, \ln{\mu^2 \over Q^2}
-   {1 \over 2} \, \left ( \left ( \gamma_{m, m}^{(0)} + \gamma_{n, n}^{(0)}   \right )^2
 +   \beta_0 \, \left ( \gamma_{m, m}^{(0)} + \gamma_{n, n}^{(0)}   \right )  \right ) \,
\ln^2 {\mu^2 \over Q^2}
\nonumber \\
&& \hspace{0.5 cm}  + \, \left ( \gamma_{m, m}^{(0)} + \gamma_{n, n}^{(0)}   \right ) \, \ln  {\mu^2 \over Q^2} \,
{\color{blue}  \left (  \beta_0 \, \ln{\nu^2 \over Q^2}
+ \left ( \gamma_{m, m}^{(0)} + \gamma_{n, n}^{(0)}   \right ) \, \ln  {\mu^2 \over Q^2}  +  {\cal R}_{m,  n}^{(1)} \right ) }
\nonumber \\
&&  \hspace{0.5 cm} -  \,  \beta_0 \, \ln{\nu^2 \over Q^2} \,  \left ( \gamma_{m, m}^{(0)} + \gamma_{n, n}^{(0)}   \right ) \,
\, \ln{\mu^2 \over Q^2}   +   {\cal R}_{m,  n}^{(2)}     \bigg  ]
+ \, {\color{magenta}{\cal O}(\alpha_s^3) }\bigg \},
\\
\nonumber
\end{eqnarray}
by inserting the Gegenbauer expansion of the  pion distribution amplitude (\ref{Gegenbauer expansion of pion LCDA})
into the hard-collinear factorization formula (\ref{factorization formula})
and then by evaluating  the double convolution integrals  over the quark momentum fractions analytically.
We can readily verify that the two coefficient matrices $\bm{\gamma}^{(0)}$ and $\bm{\gamma}^{(1)}$
coincide with the one- and two-loop anomalous dimensions (apart from  an overall minus sign)
of the flavour-octet local operators  defining the conformal moments of the twist-two distribution amplitude $\phi_\pi$.
The explicit expressions of these anomalous dimensions are   given by \cite{Mueller:1993hg}
\begin{eqnarray}
\gamma_{m, n}^{(0)}  &=&  \gamma_{m}^{(0)}   \,\, \delta_{m n}  =
C_F \, \left [ 4 \, S_1 (m+1) - \frac{2}{(m+1)(m+2)} - 3 \right ] \,\, \delta_{m n} \,,
\nonumber \\
\gamma_{m, n}^{(1)}  &=&  \gamma_{m}^{(1)} \,\, \delta_{m n}
+  \frac{(n+1)(n+2)}{(m+1)(m+2)}  \,
\frac{\gamma_{m}^{(0)}  - \gamma_{n}^{(0)}}{\bm{a} (m, n)} \,
\left [ 2 \, (2 m +3)  \left (\beta_0 + \gamma_{n}^{(0)} \right )
-  {2 m + 3 \over 2 n +3}  \bm{w}_{m, n}^{(0)}\right ] \, \vartheta_{m n}  \,,
\end{eqnarray}
where we have introduced the following definitions and conventions \cite{Braun:2017cih,Strohmaier:2018tjo,Belitsky:2005qn}
\begin{eqnarray}
S_{\ell}(m) &=&  \sum_{k=1}^{m} \, {1 \over k^{\ell}}\,,
\qquad
S_{\ell}^{\prime}(m) = 2^{\ell -1} \, \sum_{k=1}^{m} \, \left [ 1 + (-1)^{k}  \right ] \, {1 \over k^{\ell}} \,,
\qquad
\tilde{S}(m) =   \sum_{k=1}^{m} \, {(-1)^k \over k^2} \, S_1(k)  \,,
\nonumber \\
\bm{a} (m, n)  &=&  (m-n) \, (m+n +3) \,,
\qquad
\vartheta_{m n} =   \left\{
\begin{array}{l}
1 \hspace{2.0 cm}   {\rm if} \,\,  m-n>0  \,\, {\rm and \,\, even} \\
0  \hspace{2.0 cm}  {\rm else} \,.
\end{array}
\right.
\end{eqnarray}
It is straightforward to express the diagonal two-loop anomalous dimensions  in terms of the preceding  harmonic sums
\begin{eqnarray}
\gamma_{m}^{(1)} &=& 2 \, \left (C_F^2 - {1 \over 2} \, C_F \, C_A \right ) \,
\bigg  [   \frac{4 \, (2 m+3)}{ (m+1)^2 (m+2)^2} \, S_1(m+1)
-2 \, \frac{3 \, m^3 + 10 \, m^2 + 11 \, m +3}{(m+1)^3 (m+2)^3}
\nonumber \\
&& + \, 4 \, \left ( 2 \, S_1 (m+1) - {1 \over (m+1)(m+2)}  \right ) \,
\left ( S_2(m+1)  - S_2^{\prime}(m+1) \right )
+ 16 \, \tilde{S}(m+1)
\nonumber \\
&&  + 6\, S_2(m+1)  - {3 \over 4} -2 \, S_3^{\prime}(m+1)
- 4 \, (-1)^{m+1} \, \frac{2 \, m^2 + 6 \, m +5} {(m+1)^3 (m+2)^3} \bigg ]
\nonumber \\
&& + \, 2 \, C_F \, C_A \,  \bigg [S_1 (m+1) \, \left ( {134 \over 9}
+ \frac{2 \, (2 m+3)}{(m+1)^2 (m+2)^2} \right )
- 4 \, S_1(m+1) \, S_2(m+1)
\nonumber \\
&& + \, S_2(m+1) \, \left ( - {13 \over 3} + \frac{2}{(m+1)(m+2)} \right )
- {43 \over 24}
- {1 \over 9} \, \frac{151 \, m^4 + 867 \, m^3 + 1792 \, m^2 + 1590 \, m + 523} {(m+1)^3 (m+2)^3}  \bigg ]
\nonumber \\
&& + \, 2 \, C_F \, n_{\ell}  \, T_F \,  \left  [ - {40 \over 9} \, S_{1}(m+1)
+ {8 \over 3} \, S_{2}(m+1)  + {1 \over 3} \,
+ {4 \over 9} \, \frac{11 \, m^2 + 27 \, m +13}{(m+1)^2 (m+2)^2}  \right ].
\end{eqnarray}
The non-trivial  matrix $\bm{w}^{(0)}$ arises from the one-loop conformal anomaly \cite{Braun:2017cih}
\begin{eqnarray}
\bm{w}_{m n}^{(0)}  &=&  4 \, C_F \, (2 n+3) \,\bm{a} (m, n) \,
\left [ \frac{A_{m n} - S_1(m+1)}{(n+1)(n+2)}
+ { 2 \, A_{m n} \over \bm{a} (m, n)}   \right ],
\end{eqnarray}
where
\begin{eqnarray}
A_{m n} = S_1 \left ( {m+n+2 \over 2}  \right )  - S_1 \left ( {m - n - 2 \over 2}  \right )
+2 \, S_1(m-n-1)  - S_1(m+1) \,.
\end{eqnarray}
We now turn to present the desired expressions for the scattering kernels ${\cal R}^{(1)}$ and ${\cal R}^{(2)}$
by truncating the Gegenbauer expansion (\ref{Gegenbauer expansion of pion LCDA}) at $m=12$
(which is sufficient for practical purposes)
\begin{align}
\mathcal{R}_{0, 0}^{(1)} & = {79 \over 3},
& \mathcal{R}_{2, 0}^{(1)} & = {2959 \over 54} - 4 \, \zeta_3,
\nonumber \\
\mathcal{R}_{2, 2}^{(1)} & = {20563 \over 252} + {24 \over 7} \, \zeta_3,
& \mathcal{R}_{4, 0}^{(1)} & = {47537 \over 675} - 4 \, \zeta_3,
\nonumber \\
\mathcal{R}_{4, 2}^{(1)} & = {732221 \over 5400} -  24 \, \zeta_3,
& \mathcal{R}_{4, 4}^{(1)} & = {674701  \over 7425} + {360 \over 11} \, \zeta_3,
\nonumber \\
\mathcal{R}_{6, 0}^{(1)} & = {130661509 \over 1587600} -  4 \, \zeta_3,
& \mathcal{R}_{6, 2}^{(1)} & = {239823149  \over 1587600} - 24 \, \zeta_3,
\nonumber \\
\mathcal{R}_{6, 4}^{(1)} & = {348346949 \over 1587600} -  60 \, \zeta_3,
& \mathcal{R}_{6, 6}^{(1)} & = {244715911  \over 7938000} + 112 \, \zeta_3,
\\
\nonumber \\
\mathcal{R}_{8, 0}^{(1)} & = {109546973 \over 1190700} -  4 \, \zeta_3,
& \mathcal{R}_{8, 2}^{(1)} & = {3114185693  \over 19051200} - 24 \, \zeta_3,
\nonumber \\
\mathcal{R}_{8, 4}^{(1)} & = {4439757281 \over 19051200} -  60 \, \zeta_3,
& \mathcal{R}_{8, 6}^{(1)} & = {29960812963  \over 95256000} - 112 \, \zeta_3,
\nonumber \\
\mathcal{R}_{8, 8}^{(1)} & = -{7475238623 \over 60328800} + {5040 \over 19} \, \zeta_3,
& \mathcal{R}_{10, 0}^{(1)} & = {77040022223  \over 768398400} - 4 \, \zeta_3,
\nonumber \\
\mathcal{R}_{10, 2}^{(1)} & = {1002154859077 \over 5762988000} - 24 \, \zeta_3,
& \mathcal{R}_{10, 4}^{(1)} & = {352200966463  \over 1440747000} - 60 \, \zeta_3,
\nonumber \\
\mathcal{R}_{10, 6}^{(1)} & = {209131003489 \over 640332000} - 112  \, \zeta_3,
& \mathcal{R}_{10, 8}^{(1)} & = {11399947544291  \over 26893944000} - 180 \, \zeta_3,
\nonumber \\
\mathcal{R}_{10, 10}^{(1)} & = - {61920191868971 \over 154640178000} + {11880 \over 23}  \, \zeta_3,
& \mathcal{R}_{12, 0}^{(1)} & = {125618726214131  \over 1168733966400} - 4 \, \zeta_3,
\nonumber \\
\mathcal{R}_{12, 2}^{(1)} & =  {267253043878417 \over 1460917458000} - 24  \, \zeta_3,
& \mathcal{R}_{12, 4}^{(1)} & = {743080051414439  \over 2921834916000} - 60 \, \zeta_3,
\nonumber \\
\mathcal{R}_{12, 6}^{(1)} & =  {1969363800385939 \over 5843669832000} - 112  \, \zeta_3,
& \mathcal{R}_{12, 8}^{(1)} & = {2963676052586981  \over 6817614804000} - 180 \, \zeta_3,
\nonumber \\
\mathcal{R}_{12, 10}^{(1)} & =  {22456093016133983 \over 40905688824000} - 264  \, \zeta_3,
& \mathcal{R}_{12, 12}^{(1)} & = - {65202266056847263  \over 78889542732000} + {8008 \over 9} \, \zeta_3,
\end{align}
and
\begin{align}
\mathcal{R}_{0, 0}^{(2)} & =   
{13136 \over 9} - {1100 \over 9} \, \zeta_2 - {1736 \over 3} \, \zeta_3
- 24 \, \zeta_4 + {3280 \over 9} \, \zeta_5,
\nonumber \\
\mathcal{R}_{2, 0}^{(2)} & = 
 {254500067 \over 58320} - {23684 \over 81} \, \zeta_2 - {429964 \over 405} \, \zeta_3
- {1072 \over 27} \, \zeta_4 + {1240 \over 9} \, \zeta_5 - {64 \over 3} \, \zeta_2 \, \zeta_3,
\nonumber \\
\mathcal{R}_{2, 2}^{(2)} & = 
-  {977775959 \over 2857680} - {148996 \over 567} \, \zeta_2 - {163637624 \over 6615} \, \zeta_3
- {464 \over 63} \, \zeta_4 + {750200 \over 21} \, \zeta_5 + {128 \over 7} \, \zeta_2 \, \zeta_3,
\nonumber \\
\mathcal{R}_{4, 0}^{(2)} & = 
{257785535129 \over 40824000} - {740213 \over 2025} \, \zeta_2 - {76336859 \over 56700} \, \zeta_3
- {5588 \over 135} \, \zeta_4 + {1420 \over 9} \, \zeta_5 - {64 \over 3} \, \zeta_2 \, \zeta_3,
\nonumber \\
\mathcal{R}_{4, 2}^{(2)} & = 
{1065085696391 \over 61236000} - {2216813 \over 2025} \, \zeta_2 - {25432877\over 4725} \, \zeta_3
- {6776 \over 45} \, \zeta_4 + {2440 \over 3} \, \zeta_5 - 128 \, \zeta_2 \, \zeta_3,
\nonumber \\
\mathcal{R}_{4, 4}^{(2)} & = 
-  {11028087335760793 \over 148191120000} + {4090648 \over 7425} \, \zeta_2 - {8897578969 \over 38115} \, \zeta_3
+ {5360 \over 33} \, \zeta_4 + {3894200 \over 11} \, \zeta_5 + {1920 \over 11} \, \zeta_2 \, \zeta_3,
\nonumber \\
\mathcal{R}_{6, 0}^{(2)} & = 
{836512595865907 \over 105019740000} - {500357213 \over 1190700} \, \zeta_2 - {307464947 \over 198450} \, \zeta_3
- {13394 \over 315} \, \zeta_4 + {560 \over 3} \, \zeta_5  -  {64 \over 3} \, \zeta_2 \, \zeta_3,
\nonumber \\
\mathcal{R}_{6, 2}^{(2)} & = 
{301245598893433619 \over 13862605680000} - {469249769 \over 396900} \, \zeta_2
- {14861306149 \over 2182950} \, \zeta_3 - {49564 \over 315} \, \zeta_4 + {7840  \over 9} \, \zeta_5
-  128 \, \zeta_2 \, \zeta_3,
\nonumber \\
\mathcal{R}_{6, 4}^{(2)} & = 
{404162543588361839 \over 11263367115000} - {7248599 \over 2940} \, \zeta_2
- {75906747509 \over 5675670} \, \zeta_3 - {24110 \over 63} \, \zeta_4 + {17920 \over 9} \, \zeta_5
-  320 \, \zeta_2 \, \zeta_3,
\nonumber \\
\mathcal{R}_{6, 6}^{(2)} & = 
-  {1225342905440097358661 \over 3003564564000000} + {10130732267 \over 2976750} \, \zeta_2
- {2240710975609 \over 2027025} \, \zeta_3 + {30472 \over 45} \, \zeta_4 + {15187760 \over 9} \, \zeta_5
\nonumber \\
& \hspace{0.5 cm} +  {1792 \over 3}  \, \zeta_2 \, \zeta_3,
  \\
\nonumber \\
\mathcal{R}_{8, 0}^{(2)} & = 
{37633208029413760787 \over 3992430435840000} - {110645531 \over 238140} \, \zeta_2
- {44934684233 \over 26195400} \, \zeta_3 - {24592 \over 567} \, \zeta_4 + {2020 \over 9} \, \zeta_5
-  {64 \over 3}  \, \zeta_2 \, \zeta_3,
\nonumber \\
\mathcal{R}_{8, 2}^{(2)} & = 
{163112130688903608307 \over 6487699458240000} - {178296143 \over 142884} \, \zeta_2
- {3081644849093 \over 397296900} \, \zeta_3 - {30704 \over 189} \, \zeta_4 + {2840 \over 3} \, \zeta_5
-  128 \, \zeta_2 \, \zeta_3,
\nonumber \\
\mathcal{R}_{8, 4}^{(2)} & = 
{2254739676926391897989 \over 51901595665920000} - {4575297001 \over 1786050} \, \zeta_2
- {781344803377 \over 45405360} \, \zeta_3 - {74744 \over 189} \, \zeta_4 + {6200 \over 3} \, \zeta_5
-  320  \, \zeta_2 \, \zeta_3,
\nonumber \\
\mathcal{R}_{8, 6}^{(2)} & = 
{481102607081572963427 \over 8109624322800000} - {53979658807 \over 11907000} \, \zeta_2
- {1404431436769 \over 56756700} \, \zeta_3 - {303344 \over 405} \, \zeta_4 + {33160 \over 9} \, \zeta_5
-  {1792 \over 3}  \, \zeta_2 \, \zeta_3,
\nonumber   \\
\mathcal{R}_{8, 8}^{(2)} & = 
-  {180398275019354202293598809 \over 131155332247779840000} + {323163339253 \over 33934950} \, \zeta_2
- {14833184489256277 \over 4097833740} \, \zeta_3 + {299176 \over 171} \, \zeta_4 + {105065000 \over 19} \, \zeta_5
\nonumber \\
& \hspace{0.5 cm} + {26880 \over 19}  \, \zeta_2 \, \zeta_3,
\nonumber \\
\mathcal{R}_{10, 0}^{(2)} & = 
{1856383457381690687076503 \over 172702559578348800000} - {72378211243 \over 144074700} \, \zeta_2
- {697029440729 \over 374594220} \, \zeta_3 - {1373924\over 31185} \, \zeta_4 + {2440 \over 9} \, \zeta_5
-  {64 \over 3}  \, \zeta_2 \, \zeta_3,
\nonumber \\
\mathcal{R}_{10, 2}^{(2)} & = 
{606035400964911136878473 \over  21587819947293600000} - {2810106246419 \over 2161120500} \, \zeta_2
- {14893752681371 \over 1748106360} \, \zeta_3 - {1731448 \over 10395} \, \zeta_4 + 1040 \, \zeta_5
-  128  \, \zeta_2 \, \zeta_3,
\nonumber \\
\mathcal{R}_{10, 4}^{(2)} & = {2105883569365437834152149 \over  43175639894587200000} - {5694381908423 \over 2161120500} \, \zeta_2
- {90117596993 \over 4624620} \, \zeta_3 - {843548 \over 2079} \, \zeta_4 + 2160\, \zeta_5
-  320  \, \zeta_2 \, \zeta_3,
\nonumber \\
\mathcal{R}_{10, 6}^{(2)} & = {2014029806290297262575518833 \over  27891463371903331200000} - {10031992103341 \over 2161120500} \, \zeta_2
- {33180462838087841 \over 1008282775500} \, \zeta_3 - {684448 \over 891} \, \zeta_4 + {34000 \over 9} \, \zeta_5
\nonumber \\
& \hspace{0.5 cm}  - {1792 \over 3}  \, \zeta_2 \, \zeta_3,
\nonumber \\
\mathcal{R}_{10, 8}^{(2)} & = {1060110794989443086553539773 \over  12202515225207707400000} - {55755464247419 \over 7563921750} \, \zeta_2
- {23461093286572259 \over 604969665300} \, \zeta_3 - {290672 \over 231} \, \zeta_4 + {17680 \over 3} \, \zeta_5
\nonumber \\
& \hspace{0.5 cm}  -  960  \, \zeta_2 \, \zeta_3,
\nonumber \\
\mathcal{R}_{10, 10}^{(2)} & = -{7623163052247813164197161161 \over  2113433554775142600000}
+  {3521098742449999 \over 173970200250} \, \zeta_2
- {4038145298287288999 \over 427846193775} \, \zeta_3 +  {1741904 \over 483} \, \zeta_4
\nonumber \\
& \hspace{0.5 cm}  + {331478840 \over 23} \, \zeta_5 + {63360 \over 23}   \, \zeta_2 \, \zeta_3,
\nonumber \\
\mathcal{R}_{12, 0}^{(2)} & =  {27238692085662585312666509011 \over  2276565140361793881600000}
-  {58662551981653 \over 109568809350} \, \zeta_2
- {3328917505379 \over 1669619952} \, \zeta_3 -  {18093512 \over 405405} \, \zeta_4 + {980 \over 3} \, \zeta_5
\nonumber \\
& \hspace{0.5 cm}   -  {64 \over 3}   \, \zeta_2 \, \zeta_3,
\nonumber \\
\mathcal{R}_{12, 2}^{(2)} & =  {9275362166959330623335552267 \over  302356307704300749900000}
-  {1472748757823173 \over 1095688093500} \, \zeta_2
- {34886708137087 \over 3801366855} \, \zeta_3 -  {22973824 \over 135135} \, \zeta_4 + {10360 \over 9} \, \zeta_5
\nonumber \\
& \hspace{0.5 cm}   -  128  \, \zeta_2 \, \zeta_3,
\nonumber \\
\mathcal{R}_{12, 4}^{(2)} & =  {39131838489042454388302002242581 \over  735330540336859423756800000}
-  {2951046406156861 \over 1095688093500} \, \zeta_2
- {10743538046061211 \over 505581791715} \, \zeta_3 -  {11198624 \over 27027} \, \zeta_4
\nonumber \\
& \hspace{0.5 cm}   + {20440 \over 9} \, \zeta_5 -  320  \, \zeta_2 \, \zeta_3,
\nonumber \\
\mathcal{R}_{12, 6}^{(2)} & =  {7370624376245151247754283251569 \over  91916317542107427969600000}
-  {1479636772346807 \over 313053741000} \, \zeta_2
- {628083708662162821 \over 16852726390500} \, \zeta_3 -  {9083824 \over 11583} \, \zeta_4
\nonumber \\
& \hspace{0.5 cm}   + {35000 \over 9} \, \zeta_5 -  {1792 \over 3}  \, \zeta_2 \, \zeta_3,
\nonumber \\
\mathcal{R}_{12, 8}^{(2)} & =  {2135300631885046576864376162630941 \over  19731369499039061204140800000}
-  {14371610356485754 \over 1917454163625} \, \zeta_2
- {29274971196758882773 \over 542657789774100} \, \zeta_3 -  {1285412 \over 1001} \, \zeta_4
\nonumber \\
& \hspace{0.5 cm}   + {54040 \over 9} \, \zeta_5 -  960  \, \zeta_2 \, \zeta_3,
\nonumber \\
\mathcal{R}_{12, 10}^{(2)} & =  {34759246718509381395382960927377323 \over  295970542485585918062112000000}
-  {84601386041751247 \over 7669816654500} \, \zeta_2
- {73968520281383371781 \over 1356644474435250} \, \zeta_3
\nonumber \\
& \hspace{0.5 cm}   -  {23555288 \over 12285} \, \zeta_4 + {77560 \over 9} \, \zeta_5 - 1408  \, \zeta_2 \, \zeta_3,
\nonumber \\
\mathcal{R}_{12, 12}^{(2)} & =  -{202338402125338680136544777448677770453 \over  25115214605205433618413504000000}
+ {6748656400305647 \over 182614682250} \, \zeta_2
- {2829435642800818984903 \over 134173629339750} \, \zeta_3
\nonumber \\
& \hspace{0.5 cm}   +  {23637968 \over 3645} \, \zeta_4 + {868886200 \over 27} \, \zeta_5 + {128128 \over 27}  \, \zeta_2 \, \zeta_3.
\end{align}
It remains important to point out the interesting relations  ${\cal R}_{m, n}^{(1)} = {\cal R}_{n, m}^{(1)}$
and ${\cal R}_{m, n}^{(2)}={\cal R}_{n, m}^{(2)}$ on account of the charge-conjugation symmetry of
the pion form factor.

\end{widetext}

\bibliographystyle{apsrev4-1}

\bibliography{References}

\end{document}